\documentclass{epl}

\usepackage{amsmath}
\usepackage{amssymb}
\usepackage{graphicx}
\DeclareGraphicsExtensions{.eps}

\title{Length-dependent oscillations of the conductance through atomic
       chains: The importance of electronic correlations} 
\shorttitle{Oscillations in the conductance of correlated wires}

\author{
Rafael A. Molina\inst{1} 
\and Dietmar Weinmann\inst{2} 
\and Jean-Louis Pichard\inst{1,3}
}
\shortauthor{R.A.\ Molina \etal}

\institute{
\inst{1} CEA/DSM, Service de Physique de l'Etat Condens{\'e},
Centre d'Etudes de Saclay, 91191 Gif-sur-Yvette, France \\
\inst{2} Institut de Physique et Chimie des Mat{\'e}riaux de Strasbourg,
UMR 7504 (CNRS-ULP), 23 rue du Loess, BP 43, 67034 Strasbourg Cedex 2,
France \\
\inst{3} Laboratoire de Physique Th\'eorique et Mod\'elisation, 
Universit\'e de Cergy-Pontoise, 95031 Cergy-Pontoise Cedex, France
}


\begin{document}

\maketitle

\begin{abstract}
We calculate the conductance of atomic chains as a function of their 
length. Using the Density Matrix Renormalization Group algorithm 
for a many-body model which takes into account electron-electron 
interactions and the shape of the contacts between the chain and 
the leads, we show that length-dependent oscillations of the 
conductance whose period depends on the electron density in the 
chain can result from electron-electron scattering alone. The 
amplitude of these oscillations can increase with the length of 
the chain, in contrast to the result from approaches which 
neglect the interactions. 
\end{abstract}


Quantum wires of atomic dimension are among the basic components of 
molecular electronics. The knowledge of the transport
properties of such structures will be crucial for the design of
devices, and the understanding of their properties is a major
challenge in nanoelectronics. Experimentally, chains of atoms between 
two electrodes have 
been formed and investigated \cite{Ohnishi98,Yanson98,agrait_review}, 
and the conductance of molecules like H$_2$ has been measured \cite{H2}. 
Besides the possible applications in molecular devices, 
this is of great general interest since electronic correlations alter 
considerably the properties of one-dimensional systems. 
Fermi liquid theory completely looses its applicability and 
Luttinger liquid \cite{Luttinger} behavior appears. 
This has been observed for the case of metallic single wall carbon 
nanotubes \cite{carbonluttinger}.
A recent measurement on atomic Au, Pt and Ir wires between break
junction contacts \cite{Smit03} has revealed oscillations 
of the conductance as a function of the length of the chain.  
For an odd number of atoms in the atomic wire the conductance has 
maxima while it is reduced when the number of
atoms is even. This behavior is obtained after averaging over many 
realizations of the experiment in which the precise form of the contacts 
between the electrodes and the atomic chain might be different.

Since a few years, the conductance of chains of atoms has been
the subject of an increasing theoretical activity 
(see Sect.~11.5.1 of Ref.~\cite{agrait_review} for an overview),
and parity-oscillations had been expected for chains of monovalent
atoms like Au. The experimental observation of even-odd oscillations 
in chains of the polyvalent atoms Pt and Ir \cite{Smit03} however 
calls for more profound studies. 

For a non-interacting half-filled tight-binding chain coupled to 
leads, Fabry-Perot like interferences lead to size-dependent 
parity oscillations in the conductance \cite{pernas90}.
Taking into account the electron-electron interactions 
perturbatively \cite{Oguri} within the Hubbard model and
using density functional theory for chains of Na atoms \cite{sim01}
indicate that these parity oscillations of the conductance persist 
once interactions are included in the model. 
Local charge neutrality has been identified
\cite{sim01,havu02,claro} as the main cause 
of the even-odd oscillations in chains of monovalent atoms with a
half-filled valence band. However, the phase
of the parity oscillations depends on details of the model used to
describe the coupling of the chain to the leads \cite{havu02}. 
A four-atom period was obtained from density functional 
calculations for the conductance of atomic Al chains
\cite{thygesen03}. 
The analysis of the phase of these oscillations lead to the 
conclusion that the electron density in the chain is determined 
by the Fermi energy in the leads, and that 
local charge neutrality is not satisfied inside the chains. 

Taking into account the full electronic correlations 
within a model of spinless fermions with next-neighbor interactions
at half filling we showed that the even-odd oscillations can simply 
be the result of the presence of strong electronic correlations 
\cite{molin03}. We even found that the interactions could enhance the 
amplitude of the parity oscillations in the conductance.

In this work we present a study of the length dependent 
oscillations in the conductance of atomic wires, taking fully into 
account the electronic correlations inside the chain. 
After presenting our numerical results for spinless correlated 
fermions in a one-dimensional chain, we will discuss the role of 
the shape of the contacts to the leads and the dependence of the
conductance-oscillations on the particle density. These results 
will be contrasted with the theoretical predictions for models which 
neglect the interactions. This allows to identify signatures of the 
electronic correlations in the length dependent conductance oscillations 
which could be observed experimentally.


To calculate the conductance of a chain where
electronic correlations may be strong, and which is coupled to leads, 
we consider the model Hamiltonian
\begin{equation}\label{eq:hamil}
H = -t\sum_{i}(c^\dagger_i c^{\phantom{\dagger}}_{i-1} + 
                    c^\dagger_{i-1}c^{\phantom{\dagger}}_i)
+ \sum_{i=2}^{L}U_i\left[n_{i-1}-V_+\right]
\left[n_{i}-V_+\right]\, .
\end{equation}
The first term describes the kinetic energy in the atomic 
chain (from site 1 to $L$) and in the leads. 
The operators $c^\dagger_i$ create spinless fermions on the sites 
$i$ of an infinite one-dimensional tight binding model, and  
the hopping element $t=1$ is taken constant and sets the energy 
scale. The second term accounts for the interactions of strength $U_i$ 
between particles on adjacent sites $i-1$ and $i$ inside the atomic
chain. Here, $n^{\phantom{\dagger}}_i= c^\dagger_{i}c^{\phantom{\dagger}}_i$ 
is the density operator and the potential $V_+$ allows to adjust the
particle density in the chain. A density $\nu=1/2$ corresponding to half 
filling is obtained by $V_+=1/2$ which leads to particle-hole symmetry. 
To get other values for the density,
$V_+$ is chosen in a self-consistent iterative way as described in 
Ref.~\cite{molin04}. Though the one-dimensional leads are of course not 
realistic, the suppression of the electronic correlations by the screening 
in real massive electrodes is nevertheless accounted for 
by the absence of interactions.
The shrinking diameter of the contacts between the leads and the chain is 
modeled by a linear increase of the interaction strength $U_i$ from zero 
to its full value $U_i=U$ over $L_\mathrm{C}$ sites \cite{molin03}.

To determine the conductance of the atomic chain in the presence of 
electronic correlations, we exploit the connection between its 
zero-temperature linear conductance and the persistent current or 
the charge stiffness of a non-interacting ring in which the interacting 
system is embedded \cite{favand,sushk1,molin03}. 
We use the density matrix renormalization group (DMRG) algorithm
\cite{schmitteckert} to obtain the charge stiffness of the model 
(\ref{eq:hamil}) after closing it to a ring. 
For a detailed discussion of the method and the verification of 
its validity see Ref.~\cite{molin04} and references therein. 

\begin{figure}
\centerline{\includegraphics[width=0.8\textwidth]{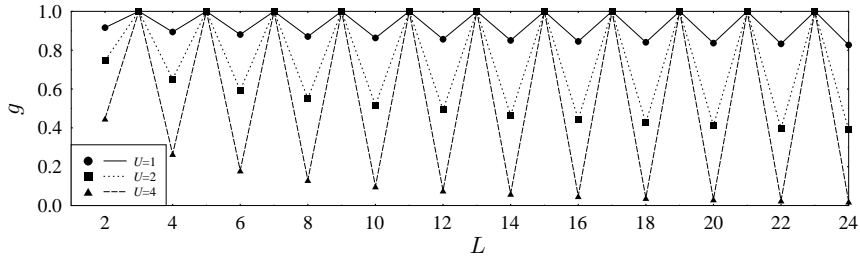}}

\caption{Conductance through interacting atomic chains at half filling 
($\nu=1/2$), as a function of their length $L$ for the case of abrupt 
contacts ($L_\mathrm{C}=1$) and different values of the interaction strength. 
}
\label{fig:evenodd}  
\end{figure}
Results for the length-dependence of the dimensionless conductance $g$ in
units of the conductance quantum $e^2/h$ are shown in
Fig.~\ref{fig:evenodd} for the case of half filling. The presence of 
interactions leads to clear parity oscillations in the conductance through 
the atomic chain, with perfect conductance $g=1$ for chains containing an odd 
number of sites, independent of the interaction strength. For even 
chains, the conductance is reduced when the interaction strength $U$ 
increases. This reduction becomes more pronounced in longer chains.

In the limit of very strong interactions, the parity oscillations at half
filling can be understood by considering the one-body hopping terms 
in (\ref{eq:hamil}) as a perturbation to the interaction term (for simplicity,
we take abrupt contacts and $U_i=U$ for all $i=\{2,\dots L\}$).
In a chain of odd length $L$, the ground state  
of the interaction term alone is a superposition of two degenerate 
components. One of them corresponds to $N=(L+1)/2$ particles occupying 
the odd sites
$i=\{1,3,\dots,L\}$, for the other $N-1$ particles occupy the even 
sites of the chain. For infinitesimal hopping 
$t/U\ll 1$, the two degenerate components become coupled through  
sequences of $N$ one-particle hopping processes. At half filling, when the 
Fermi energy in the non-interacting leads vanishes, this corresponds to a 
one-particle resonant tunneling situation. This results in perfect 
transmission and $g=1$, independent of the weakness of the coupling between 
the effective bound state and the leads which is of the order of 
$t(t/U)^{N-1}$.

For the case of an interacting chain with even length $L$, 
the ground state in the absence of the hopping term of (\ref{eq:hamil}) 
is also a superposition of two components, 
$|\Psi_\mathrm{r}\rangle>=\prod_{n=1}^N c^\dagger_{2n}|0\rangle$
and 
$|\Psi_\mathrm{l} \rangle=\prod_{n=1}^N c^\dagger_{2n-1}|0\rangle$.
They correspond to the same number $N=L/2$ of particles, occupying the even 
and the odd sites of the chain, respectively. Even though
these degenerate components are coupled by sequences of $N$ one-particle
hopping processes as in the odd case, these sequences do not involve particles
hopping between the interacting chain and the leads. Therefore, these
processes do not contribute directly to the transport through the chain.   
The process which dominates zero-temperature 
transport through the chain in the even case is coherent 
cotunneling. Taking into account the lowest order processes 
in $t/U$, involving transitions \textit{via} $N$-particle states 
with modified charge configuration and states having $N\pm 1$ 
particles in the chain leads to a decrease $\sim U^{-L}$ 
of the conductance with the interaction strength $U$ for even chain 
length $L$. 
This is a signature of the Mott insulating behavior occurring 
for infinitely long chains at $U>2$. The strong suppression 
of the even conductance with the length at $U=4$ in Fig.~\ref{fig:evenodd} 
can be understood as a precursor of the insulating behavior. 

In the limit of weak interactions, we can calculate the conductance as 
$g=4\left|G_{L,1}(0)\right|^2$,
where $G_{L,1}(0)$ is the Green's function for the propagation of a particle
from the first to the last site of the chain at the Fermi energy of
the leads $E_\mathrm{F}=0$, within the full 
many-body Hamiltonian \cite{Datta,Oguri},
evaluated at zero temperature. 
We can easily calculate this
Green's function in second order perturbation theory in $U$. 
The conductance of odd chains is not affected by this perturbation, but 
in the case of even chains it is reduced by an amount 
proportional to $U^2$, consistent with the numerical results for small 
chains and weak interaction \cite{molin03}.
Such an approach has been used to show that weakly interacting 
Hubbard chains connected to non-interacting leads
exhibit parity oscillations \cite{Oguri}.


\begin{figure}
\centerline{\includegraphics[width=0.95\textwidth]{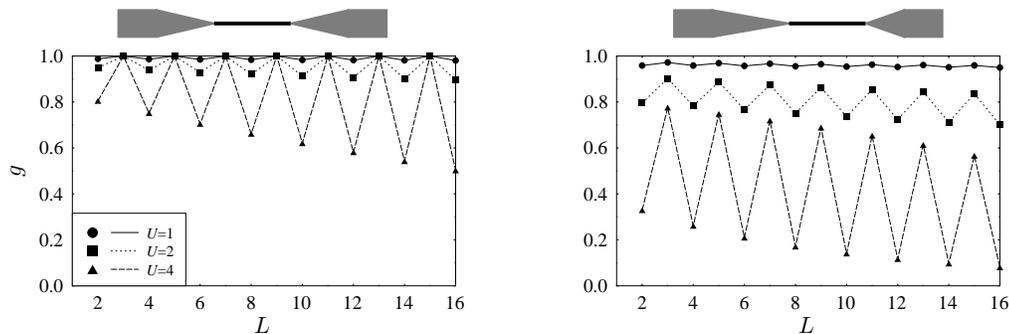}}
\caption{Conductance as a function of the number of sites in
the interacting region for $\nu=1/2$ and symmetric contacts
of length $L_\mathrm{C}=3$ (left), and asymmetric contacts of length
$L_\mathrm{Cl}=4$, $L_\mathrm{Cr}=2$ (right). The geometries we have in mind 
are sketched above the graphs.}
\label{fig:contacts}  
\end{figure}

We now address the role of the shape of the contacts between the atomic chain
and the leads in order to consider the more realistic case of smooth 
contacts. The conductance for a symmetric setup with a chain connected to 
the leads through two contacts of length $L_\mathrm{C}=3$ 
is shown in Fig.~\ref{fig:contacts} (left), at half filling. 
The even-odd oscillations as a function of the chain length 
persist in the presence of such
smooth contacts, and the perfect transmission of odd chains is not affected. 
However, the reduction of the conductance in even chains is less pronounced 
when the contacts are smooth and the amplitude of the oscillations is smaller.
Nevertheless, the dependence of the oscillations on the interaction strength 
and on the length of the chain are similar to the situation of abrupt 
contacts. Only in the unrealistic limit of infinitely smooth contacts 
one expects perfect transmission for even chains as for the odd ones 
\cite{molin03} and the parity oscillations disappear.   

While it seems reasonable to assume symmetric contacts when describing 
certain experimental situations like break junctions, it is difficult 
to exclude the case of a slightly asymmetric situation where the 
two ends of the atomic chain are connected to the electrodes in 
different ways. This breaking of
the reflection symmetry of the model has consequences that can be seen 
in Fig.~\ref{fig:contacts} (right). 
Here, the left and right contacts are assumed to be of length 
$L_\mathrm{Cl}=4$ and $L_\mathrm{Cr}=2$, respectively.
Most strikingly, the conductance obtained for odd chains no longer 
reaches the unitary limit $g=1$ and decreases with increasing
interaction strength and chain length. 
As compared to the symmetric situation, the conductance values are reduced
for the even chains as well. However, the parity oscillations persist
and the phase of the oscillations is robust as well: one has 
maxima of the conductance for odd and minima for even chains.


\begin{figure}
\centerline{\includegraphics[width=0.99\textwidth]{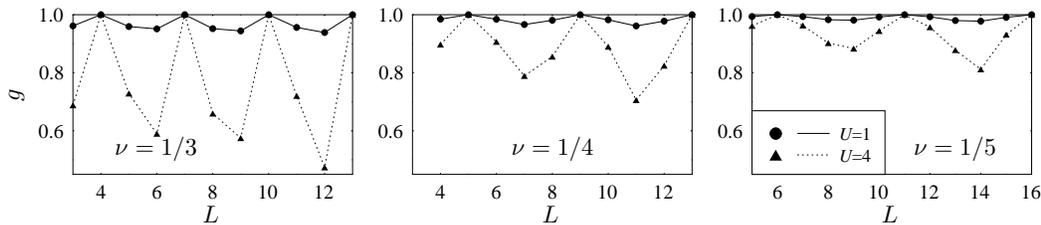}}
\caption{Conductance as a function of the length of an
atomic chain connected to the leads by abrupt contacts 
for fillings $\nu=1/3$ (left), $\nu=1/4$ (center), and $\nu=1/5$ (right).}
\label{fig:filling}  
\end{figure}
The parity oscillations that we have studied so far occur in the case of 
half filling which corresponds to chains of monovalent atoms with 
a half-filled valence band like Na and Au. 
At other values of the filling, the period of the 
length-dependent oscillations in the conductance is different.
To illustrate its dependence on the filling, we show in 
Fig.~\ref{fig:filling} results for the densities $\nu=1/3$, 1/4, and 1/5. 
For simplicity, we use abrupt and symmetric contacts. In this case, 
perfect transmission appears for chain lengths of $L=1+q/\nu$ with
integer $q=\{0,1,\dots\}$, and the period of the length-dependent oscillations
is $1/\nu$ sites for $\nu<1/2$. At larger fillings, the density of holes
$1-\nu$ is relevant for the periodicity of the conductance. 
Furthermore, it can be seen that the amplitude of the oscillations is smaller
when the filling is lower. This can be understood from the fact that the 
conductance oscillations in our model are caused by the short range 
electron-electron interactions. At low filling, this interaction becomes less 
relevant since the probability to find two particles on adjacent sites of the
chain decreases. 


Setting the interactions to zero ($U=0$) in our model leads to a perfect
homogeneous chain which exhibits perfect transmission and conductance $g=1$
for all fillings and lengths $L$, and the length-dependence of the 
conductance disappears. This shows that electronic correlations alone 
can be the origin of length-dependent conductance oscillations in atomic 
chains. In a non-interacting model, another scattering mechanism 
is necessary for reproducing the oscillating behavior 
of the conductance. A possibility is to suppose that the contacts between 
the atomic chain and the electrodes are a source of reflections. 

This can be achieved by adding the contact term
\begin{equation}
\label{eq:tightbinding}
H_\mathrm{C} = (t-v_\mathrm{l})
\left(c^\dagger_1 c^{\phantom{\dagger}}_0
+c^\dagger_0 c^{\phantom{\dagger}}_1\right)
+(t-v_\mathrm{r}) \left(c^\dagger_{L+1}c^{\phantom{\dagger}}_L
+c^\dagger_Lc^{\phantom{\dagger}}_{L+1}\right)
\end{equation}
to our Hamiltonian (\ref{eq:hamil}). This replaces the hopping matrix elements
$t$ between the chain and the left and right leads by the couplings 
$v_\mathrm{l}$ and $v_\mathrm{r}$, respectively.
Choosing $|v_\mathrm{l}|, |v_\mathrm{r}| < t$, this describes a chain of 
$L$ atoms coupled to one-dimensional leads by weak links which give
rise to Fabry-Perot like interferences. 
In this model, the particle density in the chain is the same as the one in the
leads which is determined by the Fermi energy. This situation is different
from assuming local charge neutrality as discussed in Ref.~\cite{thygesen03}. 
Using a transfer matrix formalism we can calculate the
transmission at the Fermi energy, and using Landauer's formula 
$g=|t(k_\mathrm{F})|^2$ the 
conductance 
\begin{equation}
\label{eq:enforce_filling}
g=16\sin^4k_\mathrm{F}\left|
\left(v_\mathrm{l}-v_\mathrm{l}^{-1}\right)
\left(v_\mathrm{r}-v_\mathrm{r}^{-1}\right)
e^{2ik_\mathrm{F}L}
- \left(v_\mathrm{r}e^{2ik_\mathrm{F}}-v_\mathrm{r}^{-1}\right)
\left(v_\mathrm{l}-[v_\mathrm{l}e^{2ik_\mathrm{F}}]^{-1}\right)\right|^{-2}
\end{equation}
of the non-interacting system with the Fermi wave number $k_\mathrm{F}$. 
This result is symmetric with respect to the exchange of
$v_\mathrm{r}$ and $v_\mathrm{l}$. 

\begin{figure}
\centerline{\includegraphics[width=0.99\textwidth]{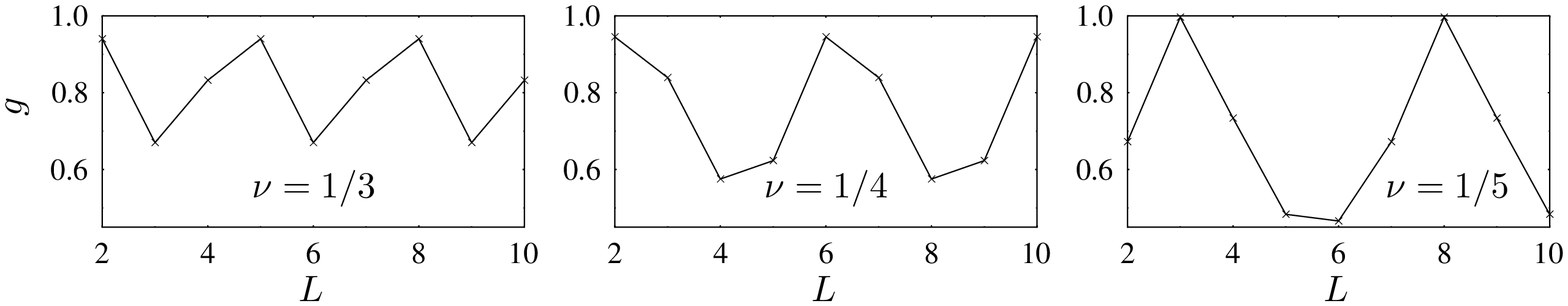}}
\caption{\label{fig:non-interacting} 
Conductance calculated from Eq.~(\ref{eq:enforce_filling}) as a function 
of the number of sites in a non-interacting chain which is connected to 
one-dimensional leads through weak links of strength
$v_\mathrm{l}=v_\mathrm{r}=0.75t$. The filling is 
$\nu=1/3$ (left), $1/4$ (center), and $1/5$ (right).} 
\end{figure}
The only $L$-dependence of $g$ in (\ref{eq:enforce_filling}) is through 
the factor $e^{2ik_\mathrm{F}L}$. Therefore, the result is periodic in $L$
with period $\pi/k_\mathrm{F}$. In dimension one, the Fermi wave
number is related to the filling through
$k_\mathrm{F}=\pi \nu$, and therefore the period of the oscillations is given 
by $1/\nu$ sites as in the interacting case discussed above.
However, the phase is different, except for half-filling, where one obtains
always even-odd oscillations with the maxima for odd $N$. 
This can be seen from Fig.~\ref{fig:non-interacting}, where the 
analytical result of Eq.~(\ref{eq:enforce_filling}) is shown for the 
example of $v_\mathrm{l}=v_\mathrm{r}=0.75$ and different fillings.
Assuming local charge neutrality and using a tight-binding model 
with a filling which is independent of the filling of the leads,
the same periods have been found in the conditions for resonant 
transport \cite{claro}. 
However, the phase of the oscillations is different \cite{thygesen03}. 

As compared to the interacting case, the behavior of the non-interacting 
model presents striking differences. First of all, without interactions 
the length-dependence is perfectly periodic and 
the conductance satisfies $g(L+1/\nu)=g(L)$. Thus, the amplitude of the
oscillations is independent of $L$. This is because the motion of 
the particles between the contacts is ballistic. 
In contrast, for interacting particles, the length-dependent oscillations 
we obtained (see Fig.~\ref{fig:evenodd}) are superimposed with a 
decrease of the conductance minima and an increase of the amplitude of 
the oscillations with the length of the chain. 
For asymmetric contacts,
the amplitude remains more or less constant, but the conductance decreases
with the length for both, maxima and minima of the oscillations 
(see Fig.~\ref{fig:contacts}). 
While such a decrease is not seen in the experimental data for 
Au chains, Pt and Ir do show \cite{Smit03} such a behavior which is 
reminiscent of our results for correlated chains, but difficult to 
explain within non-interacting models.

In addition, while the maxima in our correlated model 
always reach perfect conductance $g=1$ in the case of symmetric coupling to
the leads, without interactions this is the case
at half filling and only for special parameter combinations outside
half filling \cite{claro}.
In the case of asymmetric contacts $v_\mathrm{l}\neq v_\mathrm{r}$ the
amplitudes of the length-dependent oscillations obtained from the conductance
of Eq.~\ref{eq:enforce_filling} are somewhat reduced,  
but remain clearly observable, similar to the case of weakly correlated chains
in Fig.~\ref{fig:contacts}. This is consistent with the fact that 
the oscillations have been observed experimentally \cite{Smit03} in 
a situation where the contacts were very probably not exactly symmetric.

The period of the length-dependent oscillations predicted by both 
non-interacting approaches agrees with the result of our numerical 
calculations for a correlated chain. This indicates that the period is indeed
determined by the filling of the conductance band alone, and given by $1/\nu$,
independent of details like the shape of the contacts between the 
chain and the leads. However, outside half filling, the phase of the
oscillations is different for all three mentioned approaches. Therefore, we
can expect that, in contrast to the period, the phase depends on fine
details of the experimental situation. As a consequence, a more refined 
analysis than averaging over many realizations of the 
preparation of the chain as done in Ref.~\cite{Smit03} could be 
necessary for their observation. 

In conclusion, we have studied the effect of electronic 
correlations on the length-dependent oscillations observed in atomic 
wires. In the absence of interactions, these oscillations 
can be understood as an interference effect between two scatterers 
connecting the chain to the leads. However, the electrons of an 
atomic chain form a correlated system which can by itself be
an important source of scattering, and hence of resistance. 
We have studied the consequences of this additional scattering 
mechanism within a non-perturbative many-body approach, and have shown 
that it leads to very similar conductance oscillations. This suggests 
that electronic correlations are an additional and independent reason for 
their occurrence. In the presence of 
reflection symmetry, this mechanism leads to perfect conductance at the 
maxima of the oscillations, for all fillings. The length-dependent 
oscillations persist in the presence of asymmetric contacts between the 
chain and the leads. The period of the oscillations is given by the 
inverse of the conductance band filling, independently 
of the model and the mechanism leading to the oscillations.
But the phase of the oscillations depends on details of the model 
and whether or not interactions are present.
In non-interacting models, the electrons are mainly scattered in the
contacts, and the conductance oscillations are periodic. Electronic 
correlations lead to the occurrence of scattering processes in 
the whole chain and result in a decrease of the conductance 
with the length, in addition to the oscillations. This is a striking 
difference between our correlated chain and non-interacting models. 
As a consequence, the amplitude of the oscillations depends on the length, 
and can even increase with the length of the chain when electronic 
correlations are present. These findings should be relevant for 
distinguishing in the interpretation of recent and future experiments on 
the length-dependence of the conductance of atomic chains and molecular 
wires the many body effects coming from electron correlations from the 
effects describable by simpler one body theories.

\begin{acknowledgments}
We thank G.-L.\ Ingold and R.A.\ Jalabert for very helpful discussions and 
P.\ Schmitteckert for his DMRG program.
RAM acknowledges financial support from the European 
Union through the Human Potential Program (contract HPRN-CT-2000-00144).  
\end{acknowledgments}

\end{document}